\documentclass[showpacs,amsmath,amssymb,twocolumn,showkeys,prd,aps,10pt]{revtex4-1}
\usepackage{amsmath,color}
\usepackage{graphicx, hyperref}
\usepackage{psfrag}
\usepackage{textcomp}

\input{tcilatex}

\newcommand{\vp}{\varphi}

\begin{document}

\title{Axisymmetric Dirac-Nambu-Goto branes on Myers-Perry black hole backgrounds}
\author{Viktor G.~Czinner}
\email{czinner.viktor@wigner.mta.hu}

\affiliation
{Centro de Matem\'atica, Universidade do Minho, Campus de Gualtar, 
4710-057 Braga, Portugal\\ and\\
HAS Wigner Research Centre for Physics, 
H-1525 Budapest, P.O.~Box 49, Hungary}

\begin{abstract}
Stationary, $D$-dimensional test branes, interacting with $N$-dimensional Myers-Perry bulk black holes, 
are investigated in arbitrary brane and bulk dimensions. The branes are asymptotically 
flat and axisymmetric around the rotation axis of the black hole with a single angular momentum.
They are also spherically symmetric in all other dimensions allowing a total of $O(1)\times O(D-2)$ group 
of symmetry. It is shown that even though this setup is the most natural extension of the spherical 
symmetric problem to the simplest rotating case in higher dimensions, the obtained solutions are not 
compatible with the spherical solutions in the sense that the latter ones are not recovered in the non-rotating limit. 
The brane configurations are qualitatively different from the spherical problem, except in the special 
case of a $3$-dimensional brane. Furthermore, a quasi-static phase transition between the topologically 
different solutions cannot be studied here, due to the lack of a general, stationary, equatorial solution.   

\end{abstract}

\maketitle

\section{Introduction}

Possible interactions between branes and black holes in higher dimensions are 
interesting and important problems in many fields of modern theoretical physics. 
One direction, which has been recently introduced by Frolov \cite{Frolov}, is a 
spherically symmetric black hole interacting with Dirac-Nambu-Goto (DNG) test branes 
\cite{DNG} in arbitrary brane and bulk dimensions. This brane - black hole system,
beyond the interest of its own, has also proven to be very useful as a toy model
for various other problems. For example, it posses striking similarities 
in its properties to the problem of topology changing and merger transitions 
between higher dimensional black solutions \cite{Kol1,Kol2,Kol3}, and also shows a self-similar behavior,   
very similar to the Choptuik critical collapse phenomenon \cite{Chop}. Furthermore, it also turned
out to be a relevant model in investigating holographic phase transitions in strongly coupled gauge
theories \cite{MMT1,MMT2}, via the gauge/gravity correspondence \cite{Mald}.     
  
Generalizations to the system, by considering small thickness corrections to the branes,
have also been studied lately by Frolov and Gorbonos \cite{FG}, and more extensively 
(also within a more general framework) by us \cite{CF,C1,C2}. The motivation for this 
extension was to consider higher order, curvature corrections to the thin brane action,
which, in the holographic dual picture, correspond to finite 't Hooft coupling corrections, 
and provide a more realistic description of the  phase transition \cite{MMT2}.   

In the present paper, as a sequel to our previous works on the subject matter \cite{CF,C1,C2}, 
we provide another generalization of the problem into a different direction. 
We investigate the brane - black hole system in the rotating case by considering 
a Myers-Perry black hole in the background with a single angular momentum. The motivation
of this work is also clear, we would like to understand the role that rotational effects   
play when a quasi-static, topology changing transition is considered in the system. The 
problem is interesting not only from the geometrical point of view, but also because it 
may provide further insights to other topology changing- or merger transition problems in higher
dimensional, classical general relativity, or to  certain holographic phase transitions in the gauge/gravity 
dual picture.

In constructing the model, we
follow the method of \cite{Frolov} as closely as possible, and define the DNG-branes with
the highest possible symmetry properties that the background allows. By this construction
the branes posses a total of $O(1)\times O(D-2)$ group of symmetry, and just as in the 
spherical case, the brane action simplifies radically, resulting the problem of an ordinary
differential equation (highly non-linear though) for the brane configurations. We present and
analyze the general solution of this problem, first analytically in far distances, and
later numerically in the near horizon region. 

As a result,  we obtain that due to the coordinate 
parametrization of the Myers-Perry metric, this rotating problem is not compatible
with the spherical results of \cite{Frolov}, in the sense that the latter ones are not 
recovered in the non-rotating limit. Although the ideal situation would be to provide
a rotating solution which is the ``corresponding'' one to the spherical problem in the
above sense, nevertheless we could not find an appropriate coordinate system in which
this could be done in a natural way as presented by Frolov in \cite{Frolov}, and 
as we also do it here. Consequently, we conclude that while the construction of the
problem is the closest possible to the spherical case, the obtained results are 
qualitatively different, except in the special case of a 3-dimensional brane. 
Furthermore, we also find that stationary equatorial solutions do not generally 
exist for arbitrary brane dimensions, except again for the case of a 3-dimensional brane,
and as a result, we cannot study the quasi-static phase transition in the geometric
setup as we did in the thickness corrected spherical case \cite{CF,C1} by following 
the method of Flachi et al.~\cite{Flachi}.

The plan of the paper is as follows. In section \ref{bm}.~we define the rotating 
brane - black hole system analogous to the spherical case. In section \ref{be}.~we 
obtain the brane equation and discuss its incompatibility with the results 
of \cite{Frolov}. In section \ref{as}., first we discuss the analytic properties
of the solutions in the near horizon region and derive unique boundary conditions
for the topologically different solutions from regularity requirements. Then,
we obtain the far distance solution in an analytic form, and analyze its 
properties. In section \ref{nr}.~we present and illustrate the numerical results
in the near horizon region, and finally in section \ref{concl}.~we draw our
conclusions. In addition, we discuss the problem of the coordinate systems in 
an appendix section.

\section{The brane - black hole model }\label{bm}

The metric of the $N$-dimensional Myers-Perry solution \cite{MP} 
in Boyer-Lindquist coordinates with a single angular momentum is given by 
\begin{align}\label{MP}
ds^2&=-\left(1-\tfrac{F}{\Sigma}\right)dt^2\\
&+\sin^2\theta\left[r^2+a^2\left(1+\tfrac{F}{\Sigma}\sin^2\theta\right)\right]d\varphi^2 \nonumber\\  
&+2a\tfrac{F}{\Sigma}\sin^2\theta dtd\varphi+\tfrac{\Sigma}{\Delta}dr^2+\Sigma d\theta^2\nonumber\\ 
&+r^2\cos^2\theta d\Omega_{N-4}^2\ ,\nonumber
\end{align}
where
\begin{align}
\Sigma & =r^{2}+a^{2}\cos ^{2}\theta , \\
\Delta & =r^{2}+a^{2}-F, \\
F& =\mu r^{5-N},
\end{align}%
and $d\Omega _{N-4}^{2}$ is the line element on an $(N-4)$-dimensional unit
sphere. The parameters $\mu$ and $a$ are related to the total mass, $M$, and
angular momentum, $J$, of the black hole as
\begin{align}
M=\frac{(N-2)A_{N-2}}{16\pi G}\mu\ , \quad J=\frac{2}{N-2}Ma,
\end{align}
where 
\begin{equation}
A_{N-2}=\frac{2\pi^{\frac{N-2}{2}}}{\Gamma (\frac{N-2}{2})} 
\end{equation}
is the area of an $(N-2)$-dimensional unit sphere $S^{N-2}$. 
For simplicity, without any loss of generality, we can fix the value 
of the mass parameter $\mu$ to 1.

Test brane configurations in an external gravitational field can be obtained 
by solving the Euler-Lagrange equation derived from the Dirac-Nambu-Goto action 
\cite{DNG}
\begin{equation}\label{action0}
S=\int d^D\zeta\sqrt{-\mbox{det}\gamma_{\mu\nu}}\ ,
\end{equation}
where 
\begin{equation}
\gamma_{\mu\nu} =g_{ab}\frac{\partial x^a}{\partial \zeta^{\mu}}
\frac{\partial x^b}{\partial \zeta^{\nu}}
\end{equation}
is the induced metric on the brane and $\zeta^{\mu}(\mu=0,\dots ,D-1)$ are 
coordinates on the brane world sheet. The brane tension does not enter 
into the brane equations, thus for simplicity it can also be put equal to $1$. 
We introduce coordinates in the bulk as
\begin{equation*}
x^a=\left\{t,r,\varphi ,\theta ,\vartheta _{1},...,\vartheta
_{N-4}\right\} ,
\end{equation*}
and it is assumed that the brane is stationary, spherically symmetric 
in the $\vartheta_{i}(i=1,\dots,n=D-3)$ dimensions, rotationally symmetric 
in the $\varphi$ coordinate, and, if $D<N-1$, its surface is chosen 
to obey the equations
\begin{equation}
\vartheta_{D-2}=\dots =\vartheta_{N-4}=\pi/2\ .
\end{equation}
With the above properties the brane world sheet allows an $O(1)\times O(D-2)$
group of symmetry, and can be completely defined by the single function 
$\theta=\theta(r)$. We shall use coordinates $\zeta^{\mu}$ on the brane as
\begin{equation*}
\zeta ^{\mu }=\left\{ t,r,\varphi ,\vartheta _{1},...,\vartheta _{n}\right\},
\end{equation*}
where $n=D-3$. With this parametrization the induced metric on the brane surface 
is given by
\begin{align}
\gamma_{\mu\nu}d\zeta ^{\mu}d\zeta ^{\nu}& =-\left( 1-\tfrac{F}{\Sigma }\right)
dt^{2} \\
&+\sin ^{2}\theta \left[r^{2}+a^{2}(1
+\tfrac{F}{\Sigma}\sin ^{2}\theta)\right]d\varphi^2\notag \\
&+2a\tfrac{F}{\Sigma}\sin ^{2}\theta dtd\varphi 
+\Sigma\left(\tfrac{1}{\Delta}+\dot{\theta}^{2}\right)dr^{2}  \notag \\
&+r^{2}\cos ^{2}\theta d\Omega _{n}^{2} ,  \notag
\end{align}%
where, and throughout the paper, over-dot denotes the derivative with respect to 
the radial coordinate, $r$. 
The Dirac-Nambu-Goto action (\ref{action0}) reduces to 
\begin{equation}\label{S}
S=2\pi\Delta tA_{n}\int \mathcal{L}dr,
\end{equation}
where $\Delta t$ is an arbitrary interval of time, $A_n$ is the area of the unit 
sphere $S^n$, the $2\pi$ factor is obtained from the integration with respect to $\varphi$, and 
the Lagrangian takes the form
\begin{equation}
\mathcal{L}=r^{n}\cos ^{n}\theta \sin \theta \sqrt{\Sigma \left( 1+\Delta
\dot{\theta}^{2}\right) }.  \label{L}
\end{equation}%

\section{The brane equation}\label{be}

Test brane configurations are solutions of the Euler-Lagrange equation
\begin{equation}
\frac{d}{dr}\left( \frac{\partial \mathcal{L}}{\partial \dot{\theta}}\right)
-\frac{\partial \mathcal{L}}{\partial \theta }=0\ ,
\end{equation}%
which for the Lagrangian (\ref{L}) reads as
\begin{equation}
\ddot{\theta}+\left(\alpha\Delta +\tfrac{\dot{\Delta}}{2}\right)\dot{\theta}^{3}
+\beta\dot{\theta}^{2}+\left(\alpha+\tfrac{\dot{\Delta}}{\Delta}\right)\dot{\theta}
+\frac{\beta}{\Delta}=0,  \label{eq}
\end{equation}%
where $\alpha $ and $\beta$ are 
\begin{eqnarray}
\alpha &=&\frac{n}{r}+\frac{r}{\Sigma },\\
\beta&=&n\tan\theta-\cot\theta+\frac{a^2\sin\theta\cos\theta}{\Sigma}.
\end{eqnarray}
The horizon of the black hole is defined as the largest solution of $\Delta=0$, 
and one can consider the non-rotating problem by taking the $a\rightarrow 0$ 
limit. 

In the case of the non-rotating limit however, one notices 
that the Euler-Lagrange equation, obtained from (\ref{eq}), is not identical 
to the one that has been obtained and analyzed by Frolov in \cite{Frolov}, and 
what we also investigated in the presence of thickness corrections in the 
spherically symmetric case \cite{CF,C1,C2}. 
After some analysis one can show that the difference stems from the 
different coordinate systems used by the Myers-Perry and  
Schwarzschild-Tangherlini solutions \cite{Tangherlini}, and which disappears in standard 
$4$-dimensions in the $a\rightarrow 0$ limit, but remains present 
in higher dimensions, even after taking the non-rotating limit. 
The detailed calculation to show this is a bit lengthy, therefore we 
present it as an Appendix at the end of the paper.   
    
As a consequence, it is very important to emphasize that the DNG-brane, 
defined as $\theta(r)$ in the previous section, is \textit{not} the 
"corresponding" brane to the one that we investigated in the Schwarzschild-Tangherlini 
case, in the sense, that it does \textit{not} reproduce the 
spherical results of \cite{Frolov} in the non-rotating limit. This is because 
the angular coordinate $\theta$, through which the brane is defined in 
the Myers-Perry metric, is different from the one (denoted with the same 
letter $\theta$) in the Schwarzschild-Tangherlini solution, even after taking 
the non-rotating limit. They correspond trivially only in 4-dimensions, where 
we are accustomed to obtain the Schwarzschild coordinates in the non-rotating
limit of the Kerr solution.

It may also worth to mention that we've been trying to find an appropriate coordinate system 
for the rotating case, where those "corresponding" branes, which would reproduce the solutions 
of \cite{Frolov} as their non-rotating limit, could be defined naturally. The problem, 
however, turns out to be 
very difficult, because in those systems where the limit in the bulk is automatic, either 
the definition of the rotationally symmetric brane is problematic, or the coordinate 
transformations involve angles from the extra dimensions of the metric, which
cannot be integrated out from the action in the simple way as we did in (\ref{S}). 
Although we believe that the problem should ultimately be resolved in one way or another, 
nevertheless, we were not able to obtain a satisfactory resolution so far.  

Accordingly, in the present paper we are focusing on those DNG-branes 
which are defined in section \ref{bm}, and are the solutions of the Euler-Lagrange
equation (\ref{eq}). The problem is, of course, interesting in its own right, being the
most naturally defined DNG-brane problem on a rotating black hole background in arbitrary
dimensions, and also the most natural extension of the spherical problem to the simplest rotating case.
However, we have to keep in mind that it is essentially different from the one, that would provide 
back the Schwarzschild-Tangherlini solution of \cite{Frolov} in the non-rotating 
limit.

\section{Asymptotic and Regularity Analysis}\label{as}

In this section we present the near horizon- and far distance 
asymptotic solutions of the brane equation. From regularity 
requirements in the near horizon region, we obtain unique boundary
conditions for the problem which will be used for the numerical solution  
in the following section. 

\subsection{Near horizon behavior}

For a brane crossing the horizon (black hole embedding case or 
supercritical branch in Frolov's terminology \cite{Frolov}), 
(\ref{eq}) has a regular singular point on the horizon, $r=r_0$. A regular 
solution at this point has the following expansion near the horizon 
\begin{equation}\label{h-expand}
\theta=\theta_0+\dot\theta_0(r-r_0)+\dots,
\end{equation}
where the regularity requirement impose the condition
\begin{equation}
\dot\theta_{0}=\left.-\frac{\beta}{\dot\Delta}\right|_{r_0}\ .
\end{equation}
Consequently, supercritical solutions are all uniquely determined by their 
boundary value $\theta_0$. 

In the Minkowski embedding (subcritical) case, the brane does not cross the 
horizon, and its surface reaches its minimal distance from the black hole at 
$r_{1}>r_{0}$, which, for symmetry reasons, occurs at $\theta =0$. A regular 
(but not smooth or even differentiable, see \cite{CF,C1,C2}) solution of 
(\ref{eq}) near this point has the asymptotic behavior
\begin{equation}
\theta =\eta \sqrt{r-r_{1}}+\sigma (r-r_{1})^{3/2}+\dots \ ,
\label{o-expand}
\end{equation}%
where the regularity requirement on the axis of rotation impose the conditions
\begin{equation}
\eta=\frac{2}{\left.\sqrt{\kappa\Delta+\frac{\dot{\Delta}}{2}}\right|_{r_1}}\ ,
\end{equation}
and
\begin{eqnarray}
\!\!\!\!\!\!\!
\sigma&=&\tfrac{16}{8-9\eta^2(\dot\Delta+2\kappa\Delta)}
\left[\tfrac{\eta}{2}\left(\kappa+\tfrac{\dot\Delta}{\Delta}\right)
-\tfrac{1}{\eta\Delta}+\tfrac{\eta^3}{4}\left(n+\tfrac{1}{3}\right.\right.\nonumber\\
&+&\left.\left.\tfrac{a^2}{a^2+r^2_1}
+\tfrac{\Delta}{2}\left[\tfrac{a^2(1+\eta^2r_1)-r^2_1}{(a^2+r^2_1)^2}-\tfrac{n}{r^2_1}\right]
+\tfrac{2\kappa\dot\Delta+\ddot\Delta}{4}\right)\right]_{r_1} 
\end{eqnarray}
with 
\[
\kappa=\frac{n}{r_{1}}+\frac{r_{1}}{a^{2}+r_{1}^{2}}\ .
\]
Hence, all subcritical solutions are also uniquely determined by the single parameter, 
$r_1$.

\subsection{Far distance solution}

Since the Myers-Perry solution is asymptotically flat, the brane
function $\theta(r)$ has to converge to 
a constant value, $\theta_{\infty}$, as $r\rightarrow\infty$. The 
explicit value of $\theta_{\infty}$ is not known for the moment
(in contrast with the spherical case where it was $\pi/2$ for all dimensions), 
rather it can be obtained by the following consideration. The far 
distance solution of (\ref{eq}) can be searched in a perturbative form
\begin{equation}
\theta(r) =\theta_{\infty}+\nu(r),
\end{equation}%
where $\nu(r)$ is a first-order small function compared to 
$\theta_{\infty}$, and we require that
\begin{equation}\label{limnu}
\lim_{r\rightarrow\infty}\nu(r)=0 . 
\end{equation}
We shall only keep the linear terms of $\nu$ in (\ref{eq}) which yields 
the asymptotic equation
\begin{eqnarray}\label{Tinfeq}
\ddot{\nu}&+&\frac{n+3}{r}\dot\nu
+\frac{1+n+n\tan^2\theta_{\infty}+\cot^2\theta_{\infty}}{r^2}\nu\\
&+&\frac{n\tan\theta_{\infty}-\cot\theta_{\infty}}{r^2}
+\frac{a^2\sin\theta_{\infty}\cos\theta_{\infty}}{r^4}=0.\nonumber
\end{eqnarray} 
The general solution of (\ref{Tinfeq}) reads as
\begin{eqnarray}\label{Tinfsol}
\nu(r)=&-&\frac{B}{1+n+A}-\frac{C}{(1-n+A)r^2}\\
&+&r^{-1-\frac{n}{2}-\frac{i}{2}\sqrt{4A-n^2}}
\left[p+p'r^{i\sqrt{4A-n^2}}\right] \nonumber
\end{eqnarray}
with
\begin{eqnarray}
A&=&n\tan^2\theta_{\infty}+\cot^2\theta_{\infty},\nonumber\\
B&=&n\tan\theta_{\infty}-\cot\theta_{\infty},\nonumber\\
C&=&a^2\sin\theta_{\infty}\cos\theta_{\infty}.\nonumber
\end{eqnarray}
Before running into the analysis of the complex powers in the
solution, we notice that the first term of (\ref{Tinfsol}) is
a constant. Thus (\ref{Tinfsol}) can only be a good solution 
of (\ref{Tinfeq}) if 
$B=0$, due to the requirement (\ref{limnu}). This implies
the asymptotic constraint
\[
 n\tan\theta_{\infty}-\cot\theta_{\infty}=0,
\]
and yields the asymptotic value
\begin{equation}\label{tinf}
 \theta_{\infty}=\arctan\left[\frac{1}{\sqrt{n}}\right].
\end{equation}
According to this result, we can conclude that for each brane dimension, $n$, the solutions have 
different asymptotic behavior, and the asymptotic value, $\theta_{\infty}$, 
coincides with the Schwarzschild-value, $\pi/2$, only in the case of 
$n=0$, that is, when the brane is 3-dimensional. 

It is interesting to note here that 3-dimensional branes tend to behave differently 
from their higher dimensional counterparts in other aspects too. In our previous 
works \cite{C1,C2}, we also found that 3-dimensional branes had exceptional analytic 
properties in the near horizon region when thickness corrections had 
been considered in the non-rotating case.
 
Another interesting feature to note is that the asymptotic value does not depend
on the rotation parameter of the black hole, it is determined solely by the number
of inner dimensions of the brane in which it is spherically symmetric.   

After deriving the value for the asymptotic constants, we can obtain the 
corresponding asymptotic solutions by plugging back $\theta_{\infty}$ into (\ref{Tinfeq}),
which results the asymptotic equation 
\begin{equation}
\ddot{\nu}+\frac{n+3}{r}\dot\nu+\frac{2(n+1)}{r^2}\nu+\frac{a^2\sqrt{n}}{(n+1)r^4}=0,
\end{equation}
or plugging it directly into (\ref{Tinfsol}), and take a bit of time with the power analysis.
Either case, the asymptotic solution takes the form 
\begin{equation}\label{nu}
\!\!\!
\nu(r)=\left\{ 
\begin{array}{cc}
\frac{p\sin[\delta(r)]+p'\cos[\delta(r)]}{r^{1+\frac{n}{2}}}  
-\frac{a^2\sqrt{n}}{2(n+1)r^2},
& \mbox{if $n\leq 4 $,}\\\\
\frac{p+p'r^{\sqrt{-\gamma}}}{r^{1+\frac{n}{2}+\frac{\sqrt{-\gamma}}{2}}}
-\frac{a^2\sqrt{n}}{2(n+1)r^2}, & \mbox{if $n\geq 5$,}
\end{array}\right.
\end{equation}
where 
\begin{eqnarray}
\delta(r)=\frac{\sqrt{\gamma}}{2}\ln(r),\quad\quad \gamma=-n^2+4n+4 .
\end{eqnarray}

It may seem, for the first sight, that branes with $n\leq 4$ dimensions 
have different far distance asymptotics than the ones with dimensions 
$n\geq 5$. The real change occurs, however, at $n=3$, as we can see it from 
the following analysis. 

In the $n=0$ case, the rotation of the black hole does not seem to affect 
the asymptotic behavior directly, and, as we mentioned earlier, this 
is  the exceptional case of the 3-dimensional brane, when the asymptotic 
value, $\theta_{\infty}$, is $\pi/2$, just as in the Schwarzschild 
problem. When $n=1$, the first term dominates the solution, because the 
second one, which is controlled by the rotation parameter, decays faster. 
In the case of $n=2$, both terms decay essentially as $r^{-2}$. In all other cases, 
starting from $n\geq 3$, the first terms in the solution decay much faster
than the second one, which results that all the branes with $D=6$ or more dimensions 
have an almost uniform convergence to the asymptotic value in the far distance region,
and this is controlled by the rotation parameter of the black hole. 

The coefficients $p$ and $p'$ in the solutions are continuous functions 
of the $\theta_0$ or $r_1$ boundary parameters that we obtained previously 
from regularity requirements in the near horizon region. On the other hand, 
because of the complicated asymptotic behavior, the interpretation of $p$ or $p'$ is not 
so clear as it was in the Schwarzschild case (being the distance of the brane 
from the asymptotic value at infinity \cite{Frolov}).

\section{Numerical results}\label{nr}

After obtaining the far distance solution of the problem in analytic
form and also deriving boundary conditions from regularity requirements 
in the near horizon region, we can consider the numerical solution of
(\ref{eq}). As it was shown earlier, the boundary value $\theta_0$, or
the radial coordinate $r_1$, uniquely determines the corresponding super- 
or subcritical solutions, respectively. The numerical solution itself 
does not require very advanced techniques, we have performed it by using the 
\texttt{Mathematica}${}^\circledR{}$ \texttt{NDSolve} function.

\begin{figure}[!ht]
\noindent\hfil\includegraphics[scale=.8]{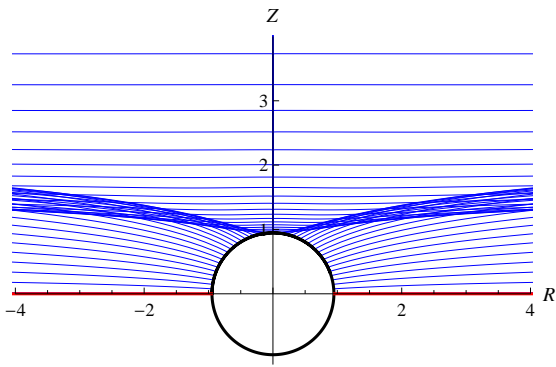} 
\caption{The picture shows a sequence of $D=3$ dimensional branes
with varying boundary values embedded in a bulk with $N=6$ dimensions.
$R$ and $Z$ are standard cylindrical coordinates, and the thick, red lines
represent the value $\theta_{\infty}$ which is $\pi/2$ for the present 
case. The value of the rotation parameter $a=0.4$.}
\label{fig:n0}
\end{figure}

On FIG.~\ref{fig:n0}~we are plotting a sequence of $D=3$ ($n=0$) brane solutions 
from both 
topologies in the near horizon region. The asymptotic constant in this special case is
$\pi/2$ and we have chosen the value $0.4$ for the rotation parameter, $a$. As a 
result (just as we expect from the far distance analysis) the brane configurations 
are very similar to what we had before in the spherical case \cite{Frolov,CF}. 

\begin{figure}[!ht]
\noindent\hfil\includegraphics[scale=.8]{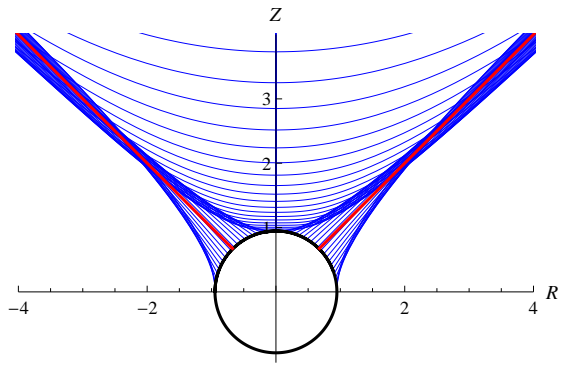} 
\caption{The picture shows a sequence of $D=4$ dimensional branes
with varying boundary values embedded in a bulk with $N=6$ dimensions.
$R$ and $Z$ are standard cylindrical coordinates, and the thick, red lines
represent the value $\theta_{\infty}$, to which the solutions asymptotically
converge, $\pi/4$ for the present case. The value of the rotation 
parameter $a=0.4$.}
\label{fig:n1}
\end{figure}

By increasing the brane dimension from $D=3$ ($n=0$) to $D=4$ ($n=1$), 
and keeping the bulk dimension fixed ($N=6$), we can see the interesting 
new result on the asymptotic behavior. We plotted this situation on 
FIG.~\ref{fig:n1}. In this case, the 
asymptotic value, $\theta_{\infty}$, is $\pi/4$, and 
it can be seen that all solution tend asymptotically to this value (in good 
agreement with the far distance analysis) independently of the near horizon 
boundary values.

\begin{figure}[!ht]
\noindent\hfil\includegraphics[scale=.8]{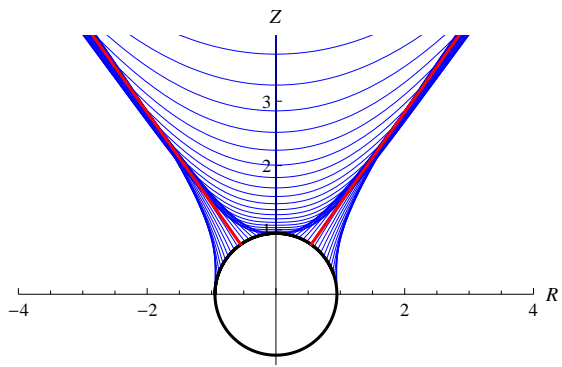} 
\caption{The picture shows a sequence of $D=5$ dimensional branes
with varying boundary values embedded in a bulk with $N=6$ dimensions.
$R$ and $Z$ are standard cylindrical coordinates, and the thick, red lines
represent the value $\theta_{\infty}$, to which the solutions asymptotically
converge, $\theta_{\infty}=\arctan[1/\sqrt{2}]$ for the present case. The value of the rotation 
parameter $a=0.4$.}
\label{fig:n2}
\end{figure}

By increasing the number of brane dimensions, $n$, the value of the asymptotic 
constant, $\theta_{\infty}$, changes according to (\ref{tinf}), but the qualitative
picture of the solutions remains essentially similar to what we see on FIG.~\ref{fig:n1}. 
For the sake of illustration, on FIG.~\ref{fig:n2}, we also plot the $D=5$ ($n=2$) dimensional case
with $N=6$.  

\begin{figure}[!ht]
\noindent\hfil\includegraphics[scale=.46]{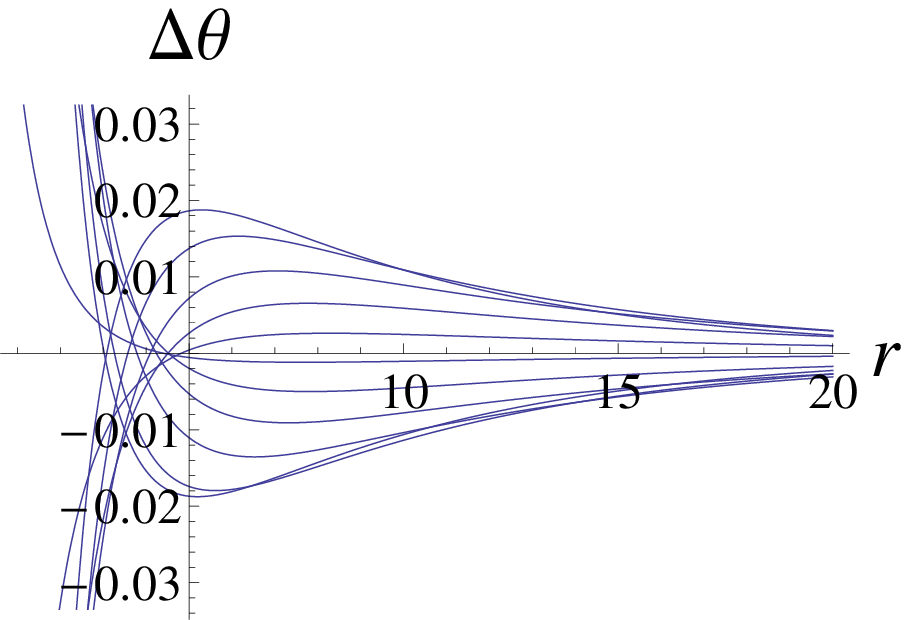} 
\noindent\hfil\includegraphics[scale=.46]{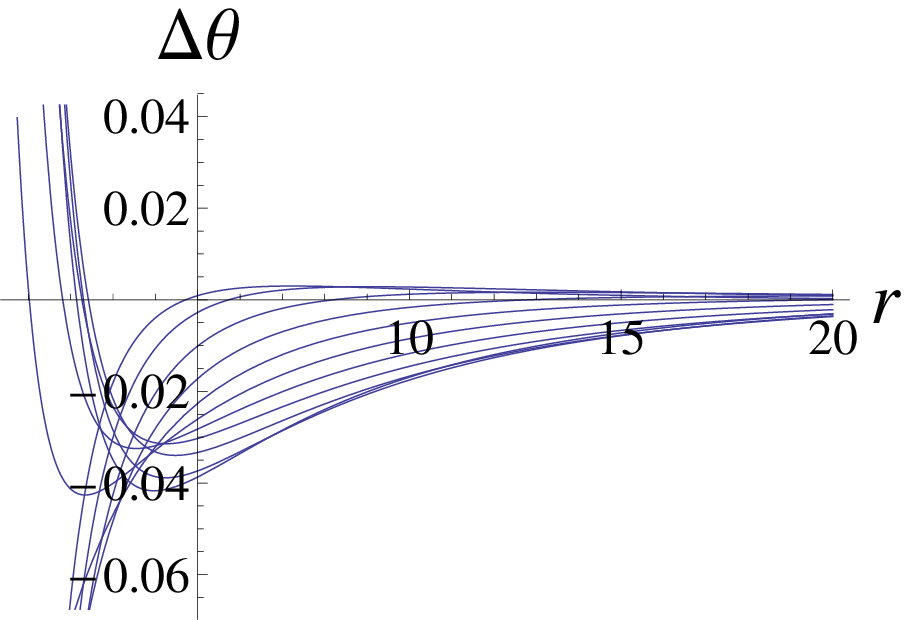}
\caption{The picture shows a sequence of $\Delta\theta(r)$ functions 
of $D=4$ dimensional branes embedded in a $N=6$ dimensional bulk. 
The boundary values are equally distributed around $\theta_{\infty}=\pi/4$ 
in the $\theta_0\in (0,\pi/2)$ region. The left picture belongs to a 
slow rotation, $a=0.1$, while the right picture belongs to the 
$a=0.9$ value.}
\label{fig:dt}
\end{figure}

By changing the value of the rotation parameter, the near horizon configurations
are also changing together with the asymptotic convergence to $\theta_{\infty}$,
that we discussed earlier in the far distance solution. In order to illustrate 
this change, we define the function
\begin{equation}
 \Delta\theta(r)=\theta(r)-\theta_{\infty},
\end{equation}
and compute the $\Delta\theta(r)$ functions for a sequence of brane solutions with different boundary
values, $\theta_0$, equally distributed around the $\theta_{\infty}=\pi/4$ value
in the $\theta_0\in (0,\pi/2)$ region, just as on FIG.~\ref{fig:n1} and FIG.~\ref{fig:n2}. 
The corresponding
curves are plotted on FIG.~\ref{fig:dt}.~for two different rotation parameter values, $a=0.1$ 
(left picture) and $a=0.9$ (right picture). 

In the case of slow rotation ($a=0.1$), the $\Delta\theta(r)$ functions have an almost
``mirror symmetric'' amplitude distribution around the $\theta_{\infty}=\pi/4$ 
value (right picture on FIG.~\ref{fig:dt}.), while in the case of a large rotation parameter ($a=0.9$),
the picture becomes very asymmetric. The branes with boundary values 
$\theta_0\in (\pi/4,\pi/2)$ deviate strongly from $\theta_{\infty}$ in the 
near horizon region, while the branes with $\theta_0\in (0,\pi/4)$ approach
the asymptotic value very quickly. 
       
In our previous works \cite{CF,C1}, we also investigated the problem of a quasi-static
evolution of a brane from the equatorial plane in black hole embeddings, 
to a Minkowski embedding topology, through a topology change transition. 
The question was very natural there,
following the method developed in \cite{Flachi}, because the equatorial configuration was
a general solution in every dimensions of the spherical problem. In the present rotating case
however, as we saw above, the equatorial configuration is a solution of the problem only in the exceptional
case of the 3-dimensional brane, and we cannot use this method for a general discussion.
Although we could analyze the topological phase transition in this special case,
nevertheless we believe that it would be misleading since the relevant problems
are usually obtained from higher dimensions, like the case of the holographic dual
phase transition. As a consequence, the question remains open in the present
rotating case.

\section{Conclusions}\label{concl}

In the present work, we studied the problem of rotationally symmetric, stationary,
Dirac-Nambu-Goto branes on the background of a Myers-Perry black hole with a single
angular momentum. In defining the interacting brane - black hole system, we strongly
followed the spherical problem given by Frolov \cite{Frolov}. Although this model 
is the most natural extension of the spherical setup to the simplest rotating case,
we found that due to the non-equivalent coordinate parametrization, the obtained
solutions are not compatible with the spherical solutions in the sense that the 
latter ones are not recovered in the non-rotating limit. Our efforts, to find an 
appropriate coordinate system in which the rotating problem could be formulated 
naturally, in a way that the spherical case could also be reproduced in the 
$a\rightarrow 0$ limit, has not succeeded so far. It is an open question
whether it can be done at all.   

After clarifying the above situation, we analyzed the properties of the obtained 
problem, and presented its solution both analytically, at far distances, and numerically, 
in the near horizon region. In the latter case, we found that the analytic properties 
of the test brane solutions, both on the axis of rotation and on the horizon, are very similar 
to what we saw in the spherical case. From regularity requirements we could obtain unique 
numerical solutions for each, freely chosen, boundary value in both topologies.      

By analyzing the  far distance solutions, we obtained a new interesting  
result that the asymptotic behavior of the rotating solutions are qualitatively
different from the spherical problem, except in the special case of a 3-dimensional brane.
This difference change the entire structure of the brane configurations in the
near horizon region too, because all solutions are attracted asymptotically to the
same constant value, independently of the near horizon boundary conditions.   
Furthermore, the asymptotic value is different for every brane dimensions, and it
is controlled solely by the dimension parameter of the brane. 

Another interesting result is that the rotation of the black hole has a direct 
effect only on how the solutions tend to the asymptotic value, and we illustrated 
this phenomenon in the cases of a small and a large rotation parameter.  

One of the motivations of this work was to understand the role that rotational 
effects may play in a quasi-static, topology changing phase transition of the system. 
As a negative result, we obtained that the problem cannot be studied here in the 
geometrical way that we applied in the thickness corrected spherical problem \cite{CF,C1}, 
due to the lack of a general, stationary, equatorial solution for arbitrary dimensions. 
Consequently, we could not obtain general results on the phase transition in this paper, 
so the question remains open for the rotating case. 

The lack of the equatorial solution has another consequence which is connected to the 
stability of the rotating brane - black hole system. It has been shown
by Hioki et al.~\cite{Hioki} that equatorial solutions are stable against small 
perturbations in the spherical case. Stability is an important 
issue in higher dimensions, 
and it would be also important to know whether similar results may hold for
the present axisymmetric case too. Unfortunately, because of the lack of the equatorial 
solution, the question of stability can not be studied here by using the method of Hioki 
et al.~for the general case.    

As another stability issue, in this paper we have not considered the cases of extremal 
and ultra-spinning black holes. The reason for 
this is the fact that ultra-spinning black holes are expected to be unstable \cite{EM},
and the instability limit occurs at a surprisingly low value of the angular 
momentum, i.e.~not far in the ultra-spinning regime. In fact, the magnitude
of the critical rotation parameter, $a$, has been estimated by Emparan and Myers
for several dimensions \cite{EM}, and it turned out that a typical value 
is around $a\approx 1.3$. According to this, in the present paper we constrained ourselves to 
keep the value of the rotation parameter small enough to stay away from the presumably 
unstable regime.

\acknowledgments
I am grateful for valuable discussions with Bal\'azs Mik\'oczi and
Alfonso Garc\'{\i}a Parrado G\'omez-Lobo. Most calculations, especially the 
numerical parts, have been performed and checked by the computer 
algebra package \texttt{MATHEMATICA 9}. The research leading to this result 
has received funding from the European Union Seventh Framework Programme 
(FP7/2007-2013) under the grant agreement No.~PCOFUND-GA-2009-246542 
and from the Foundation for Science and Technology of Portugal.

\appendix*
\section{The Schwarzschild-Tangherlini limit of the Myers-Perry solution}

The $N$-dimensional Myers-Perry metric with a single angular momentum in
Boyer-Lindquist coordinates is given in (\ref{MP}), while the 
Schwarzschild-Tangherlini (ST) solution of the same dimension is given by 
\begin{align}\label{schw}
ds^2&=-fdt^2+f^{-1}dr^2+r^2d\Omega_{N-2}^2\ ,
\end{align}
where
\begin{equation}\label{f}
f=1-\frac{\mu}{r^{N-3}}\ .
\end{equation}

In both formulas $d\Omega_k^2$ is the metric of a $k$-dimensional unit sphere $S^k$,
which is parametrized with the polar coordinates, $\xi_k$, defined by the following 
recursive relation
\begin{equation}\label{O}
d\Omega_{k+1}^2=d\xi_{k+1}^2+\sin^2\xi_{k+1}d\Omega_{k}^2\ .
\end{equation}

By taking the limit of $a\rightarrow0$ in the MP metric, the ST 
solution has to be reproduced. In order to check this, after taking 
the limit in the coefficient functions, one arrives 
to the following equation for the metric on the $(N-2)$-dimensional unit 
sphere,
\begin{equation}\label{oeq}
d\Omega_{N-2}^2=d\theta^2+\sin^2\theta d\vp^2+\cos^2\theta d\Omega_{N-4}^2\ . 
\end{equation}
Applying the recursive relation given in (\ref{O}) we can rewrite (\ref{oeq}) into the form
\begin{eqnarray}\label{oeq2}
d\xi_1^2+\sin^2\xi_1d\xi_2^2+\sin^2\xi_1\sin^2\xi_2 d\Omega_{N-4}^2=\nonumber\\
d\theta^2+\sin^2\theta d\vp^2+\cos^2\theta d\Omega_{N-4}^2\ . 
\end{eqnarray}
From (\ref{oeq2}) it is clear that if $N>4$, the angular parametrization of the 2-sphere 
in question is different from that of the ST metric of the same dimension. 
This difference however disappears in standard 4-dimensions since the last 
terms are zero on both sides yielding the equivalence 
\begin{equation}
\theta=\xi_1\ , \quad \vp=\xi_2\ . 
\end{equation}

In order to see the ST limit of the MP metric for $N>4$,
one needs to verify that the angular parametrization given in (\ref{MP}) is
 equivalent with the Schwarzschild 
parametrization. To show this, we need to find the transformation laws from the 
spherical coordinates defined by the polar angles $\xi_1$ and $\xi_2$, to the 
coordinate system defined by the angles $\theta$ and $\vp$. 
The transformation rules are the solution of the following system of equations obtained 
from (\ref{oeq2}),

\begin{eqnarray}\label{teqs}
\left(\frac{\partial\theta}{\partial\xi_1}\right)^2
+\sin^2\theta\left(\frac{\partial\vp}{\partial\xi_1}\right)^2&=&1,\\
\left(\frac{\partial\theta}{\partial\xi_2}\right)^2
+\sin^2\theta\left(\frac{\partial\vp}{\partial\xi_2}\right)^2&=&\sin^2\xi_1,\\
\frac{\partial\theta}{\partial\xi_1}\frac{\partial\theta}{\partial\xi_2}
+\sin^2\theta\frac{\partial\vp}{\partial\xi_1}\frac{\partial\vp}{\partial\xi_2}&=&0,\\
\sin^2\xi_1\sin^2\xi_2&=&\cos^2\theta,
\end{eqnarray}
where $\theta=\theta(\xi_1,\xi_2)$ and $\vp=\vp(\xi_1,\xi_2)$. This system can be integrated 
in a closed form with the solution
\begin{align}
\theta&=\arccos\left[\sin\xi_1\sin\xi_2\right],\\
\vp&=\arctan\left[\cos\xi_2\tan\xi_1\right],
\end{align}
or equivalently the inverse transformations are
\begin{align}\label{vt1}
\xi_1=\arcsin\left[\sqrt{1-\cos^2\vp\sin^2\theta}\right],\\
\xi_2=\arcsin\left[\frac{\cos\theta}{\sqrt{1-\cos^2\vp\sin^2\theta}}\right].\label{vt2}
\end{align}

To see how the angles $\theta$ and $\vp$ parametrize the unit $2$-sphere 
let us utilize the standard Cartesian coordinates $x,y,z$ given by
\begin{align}\label{polar}
x&=\sin\xi_1\cos\xi_2,\nonumber\\
y&=\sin\xi_1\sin\xi_2,\\
z&=\cos\xi_1.\nonumber
\end{align}
Here the polar angle $\xi_1\in[0,\pi]$ is measured from the positive $z$-direction, 
and the azimuthal angle $\xi_2\in[0,2\pi]$ runs in the $x-y$ plane measured from the 
positive $x$-direction. Expressing now $x$, $y$ and $z$ as functions of $\theta$ and 
$\vp$ through the transformation formulas (\ref{vt1}) and (\ref{vt2}) we get  
\begin{align}\label{xzpolar}
x&=\sin\vp\sin\theta\ ,\nonumber\\
y&=\cos\theta\ ,\\
z&=\cos\vp\sin\theta .\nonumber
\end{align}
It is thus clear that $\theta$ and $\vp$ are also spherical polar coordinates 
of the unit 2-sphere in a way that the polar angle $\theta\in[0,\pi]$ is measured from the 
positive $y$-direction, and the azimuthal angle $\vp\in[0,2\pi]$ runs in the $z-x$ plane 
measured from the positive $z$-direction. 

According to this, we observe that in standard 4 dimensions the axis of rotation of the 
Kerr black hole is orthogonal to the $x-y$ plane (corresponding to the labeling of the 
standard Cartesian coordinates above), however in dimensions $N>4$, the axis of rotation 
``switches'' to be orthogonal to the $z-x$ plane instead. One has to be thus 
very careful in taking the Schwarzschild-Tangherlini limit of the Myers-Perry solution 
in higher dimensions, because the angular parametrization of the two solutions remains 
different even in the non-rotating limit.

\end{document}